\documentclass[twocolumn]{aastex631}
\usepackage{amsmath}
\accepted{to ApJ}

\graphicspath{{fig/}}

\begin{document}
\title{Streaming Torque in Dust-Gas Coupled Protoplanetary Disks}

\correspondingauthor{Cong Yu}
\email{yucong@mail.sysu.edu.cn}

\author[0000-0002-8125-7320]{Qiang Hou}
\affiliation{School of Physics and Astronomy, Sun Yat-sen University, Zhuhai 519082, China}
\affiliation{CSST Science Center for the Guangdong-Hong Kong-Macau Greater Bay Area, Zhuhai 519082, China}
\affiliation{State Key Laboratory of Lunar and Planetary Sciences, Macau University of Science and Technology, Macau, China}
\author[0000-0003-0454-7890]{Cong Yu}
\affiliation{School of Physics and Astronomy, Sun Yat-sen University, Zhuhai 519082, China}
\affiliation{CSST Science Center for the Guangdong-Hong Kong-Macau Greater Bay Area, Zhuhai 519082, China}
\affiliation{State Key Laboratory of Lunar and Planetary Sciences, Macau University of Science and Technology, Macau, China}

\begin{abstract}
We investigate the migration of low-mass protoplanets embedded in dust-gas coupled protoplanetary disks. Linear calculations are performed with respect to the NSH (Nakagawa-Sekiya-Hayashi 1986) equilibrium within a shearing sheet. We find that the dusty quasi-drift mode dominates the dynamical behaviors in close proximity to the protoplanet. This mode exhibits an extremely short radial wavelength, characterized by a dispersion relation of $ \tilde{\omega} = \left( 1 + \mu \right) \boldsymbol{W}_s \cdot \boldsymbol{k}$. The emergence of this mode leads to a wake with a short radial length-scale ahead of protoplanets, contributing to a positive torque, termed as ``Streaming Torque (ST)''. Furthermore, both Lindblad torque and corotation torque are affected by the NSH velocity. The total torque and planetary migration are contingent upon the coupling strength between dust and gas. In most scenarios, ST predominates, inducing outward migration for planets, thereby addressing the issue of rapid inward migration in their formation paradigm.

\end{abstract}


\section{Introduction} \label{introduction}

The interactions between planets and protoplanetary disks (PPDs) play pivotal roles in the formation and evolution of planetary systems. Planetary migration within PPDs is a natural outcome of planet-disk interaction and three types of planetary migration have been extensively studied \citep{Kley2012, Paardekooper2023}. Type I migration occurs for low-mass (proto)planets \cite[e.g.][]{Goldreich1979,Goldreich1980,Ward1986,Ward1997,Korycansky1993,Tanaka2002,Yu2010,D'Angelo2010,Wu2024,Tanaka2024,Ziampras2024}. In this scenario, two components contributes: Lindblad resonance and corotation resonance. Early studies in type I migration for pure gaseous PPDs indicated rapid inward migration of planets with a timescale shorter than the lifetime of PPDs (several million years), in contradiction with both planet formation theories and observations. Additional physical ingredients such as viscosity \citep[e.g.][]{Yu2010}, self-gravity \citep[e.g.][]{Pierens2005,Baruteau2008}, magnetism \citep[e.g.][]{Terquem2003,Baruteau2011}, thermodynamics \citep[e.g.][]{Llambay2015} and gas accretion \citep{Li2024,Laune2024}  must be considered, some of which could provide angular momentum and, in some cases, lead to outward migration. However, the lingering question of fast inward migration persists when attempting to establish a self-consistent planet formation paradigm that aligns with observations of exoplanets.

Despite its small mass fraction, dust plays a significant role in both theoretical and observational studies. However, the studies of dusty torque have been very limited due to its complexity. Some studies suggest that dust could alter migration history \citep{Llambay2018,Kanagawa2019,Hsieh2020,Guilera2023}. \cite{Llambay2018} revealed the presence of a dust deficit behind the planet under specific conditions, resulting in a positive torque. The intricate dynamics near planets underscore the complexity of the phenomenon. This study aims to elucidate the fundamental dynamics of dust and gas and their implications for planetary migration, thereby shedding light on the underlying physics.

In this paper, we investigate planetary migration for a low-mass protoplanet embedded in dust-gas coupled PPDs. Section \ref{methods} introduces the basic equations and our methodology. Linear calculations are conducted within the framework of the NSH equilibrium under the shearing sheet approximation. Section \ref{results} presents our findings, revealing the emergence of a dusty ``quasi-drift" mode with a short wavelength, combined with ``quasi-density" waves, forming a multi-scale structure. We analyze their properties and morphologies. In Section \ref{torque}, we calculate torques induced by these modes with various parameters, demonstrating the possibility of planetary outward migration. Finally, we offer discussions and conclusions in Section \ref{Conclusions}.

\section{Basic Equations} \label{methods}
The hydrodynamic equations for 2D dust-gas coupled thin PPDs read
\begin{gather} 
    \frac{\partial \Sigma_d}{\partial t}+\nabla \cdot\left(\Sigma_d \boldsymbol{V}_d\right)=0,\label{govern_1} \\ 
    \frac{\partial \Sigma_g}{\partial t}+\nabla \cdot\left(\Sigma_g \boldsymbol{V}_g\right)=0, \label{govern_2}\\
    \frac{\partial \boldsymbol{V}_d}{\partial t}+\boldsymbol{V}_d \cdot \nabla \boldsymbol{V}_d= - \nabla \Phi_{\star} - \frac{\boldsymbol{V}_d-\boldsymbol{V}_g}{t_{\text {s}}}, \label{govern_3}\\
    \frac{\partial \boldsymbol{V}_g}{\partial t}+\boldsymbol{V}_g \cdot \nabla \boldsymbol{V}_g= - \nabla \Phi_{\star} +\frac{\Sigma_d}{\Sigma_g} \frac{\boldsymbol{V}_d-\boldsymbol{V}_g}{t_{\text {s}}}-\frac{\nabla P}{\Sigma_g},\label{govern_4}
\end{gather}
where $\boldsymbol{V}$, $\Sigma$, $P$ and $\Phi_{\star}$ are velocity, (surface) density, gas pressure and star potential. Subscripts ``$d$'' and ``$g$'' denote quantifies of dust and gas. $t_s$ is the stopping time. A steady solution for these equations, known as NSH (Nakagawa-Sekiya-Hayashi) equilibrium, has been given by \cite{Nakagawa1986}, i.e.
\begin{gather}
    V_{dr} = - f_g \chi_1 \eta V_{\mathrm{K}}, \label{NSH_1} \\
    V_{d\theta} = \left( 1 - f_g \chi_2 \eta \right) V_{\mathrm{K}},  \label{NSH_2} \\
    V_{gr} = f_d \chi_1 \eta V_{\mathrm{K}},  \label{NSH_3} \\
    V_{g\theta} = V_{\mathrm{K}} + \left( f_d \chi_2 - 1 \right) \eta V_{\mathrm{K}}.  \label{NSH_4} 
\end{gather}
Here $f_{d/g} \equiv \Sigma_{d/g}/( \Sigma_{d} + \Sigma_{g})$  is the mass fraction of dust/gas. $V_{\mathrm{K}} \left( \Omega_{\mathrm{K}} \right)$ represents the Keplerian speed (frequency). A useful dimensionless quantity can be defined as $ \tau \equiv \Omega_{\mathrm{K}} t_s$. Then $\chi_1 \equiv 2f_{g}\tau/ [ 1+\left( f_{g} \tau \right)^2 ]$, $\chi_2 \equiv 1/[1+ \left( f_{g} \tau \right)^2]$. And $ \eta \equiv - \frac{1}{2 \Sigma_g V_{\mathrm{K}}^2 }\frac{\partial P}{\partial \mathrm{ln} r} = 0.5 h^2 $ quantifies the radial pressure support, with the assumptions of isothermal gas and a power law dependency on radius for its density. $h$ is the aspect ratio defined as $ h \equiv \mathrm{H}/r$, where $\mathrm{H}$ is the scale height.

Equation \eqref{govern_1}-\eqref{govern_4} can be readily investigated based on the ``shearing sheet'' \citep{Goldreich1965,Narayan1987}. Linear perturbation theory is used to decompose those equations into steady and perturbation equations. A low-mass planet is embedded in the latter as a perturbation source located at the origin of the coordinate. We assume the planet is with the mass below roughly a few percent of thermal mass, $M_{\mathrm{th}}$, defined by $M_{\mathrm{th}}=h_p^{3}M_{\star}$ ($M_{\star}$ is the star mass). In this regime, linear analysis behaves well \citep{Miranda2020}, beyond which gap opening will occur and type II migration works. Subscript ``$p$'' denotes the values at the planet location. We adopt a Cartesian coordinate system with $x = \left( r - r_p \right)/\mathrm{H}$ and $y = r_p \left( \theta - \theta_p \right)/\mathrm{H}$ and neglect all curvature terms. We express perturbation quantities as integrals in Fourier space, i.e. $ X_{1}(x,y) = \int_{- \infty}^{+ \infty} \delta X(x,k_y) \exp \left( ik_y y \right) d k_y$. Lastly, we focus on the Stokes regime when dealing with drag terms, i.e. $\tau = \mathrm{const.}$, applied when the dust size is much larger than the mean free path of the gas. It has no fundamental physical difference with the Epstein regime, when we should make $\Sigma_g \tau = \mathrm{const.}$ \citep{Pan2020}. Then perturbation equations for a sepecific $k_y$ read
\begin{gather}
    \left( i k_y \frac{V_{dy}}{\Omega_p} + \frac{V_{dx}}{\Omega_p}\frac{d}{d x} \right) \frac{\delta \Sigma_d}{f_d \Sigma_{p}} + \frac{d}{d x} \frac{\delta V_{dx}}{\Omega_p} + i k_y \frac{\delta V_{dy}}{ \Omega_p} = 0, \label{linear1}\\
    \left( i k_y \frac{V_{gy}}{\Omega_p} + \frac{V_{gx}}{\Omega_p} \frac{d}{d x} \right) \frac{\delta \Sigma_g}{f_g \Sigma_{p}} + \frac{d}{d x} \frac{\delta V_{gx}}{\Omega_p} + i k_y \frac{\delta V_{gy}}{ \Omega_p} = 0, \label{linear2}\\
    \left( \frac{1}{\tau} + i k_y \frac{V_{dy}}{\Omega_p} + \frac{V_{dx}}{\Omega_p} \frac{d}{d x} \right) \frac{\delta V_{dx}}{\Omega_p} - \frac{ 2 \delta V_{dy}}{\Omega_p} - \frac{1}{\tau} \frac{\delta V_{gx}}{\Omega_p} = - \frac{1}{c_s^2} \frac{\partial \phi_p}{\partial x}, \label{linear3}\\
    \frac{\delta V_{dx}}{2 \Omega_p} + \left( \frac{1}{\tau} + i k_y \frac{V_{dy}}{\Omega_p} + \frac{V_{dx}}{\Omega_p}  \frac{d}{d x} \right) \frac{\delta V_{dy}}{\Omega_p} - \frac{1}{\tau} \frac{\delta V_{gy}}{\Omega_p} = - ik_y  \frac{ \phi_p}{c_s^2}, \label{linear4}\\
    - \frac{ W_{s,x}}{\tau} \frac{\delta \Sigma_d}{f_g \Sigma_{p}} + \left( \frac{\mu W_{s,x}}{\tau} +  \frac{d}{dx} \right) \frac{\delta \Sigma_g}{f_g \Sigma_p} - \frac{\mu }{\tau} \frac{\delta V_{dx}}{\Omega_p} \nonumber \\ 
    + \left( \frac{ \mu}{\tau} + i k_y \frac{V_{gy}}{\Omega_p} + \frac{V_{gx}}{\Omega_p} \frac{d}{dx} \right) \frac{\delta V_{gx}}{ \Omega_p}- \frac{2 \delta V_{gy}}{\Omega_p} = - \frac{1}{c_s^2} \frac{\partial \phi_p}{\partial x}, \label{linear5} \\
    - \frac{W_{s,y}}{\tau} \frac{\delta \Sigma_d}{f_g \Sigma_p} + \left( \frac{\mu W_{s,y}}{\tau}   + i k_y \right) \frac{\delta \Sigma_g}{f_g \Sigma_p}  - \frac{\mu}{\tau} \frac{\delta V_{dy}}{\Omega_p} \nonumber \\
    +  \frac{\delta V_{gx}}{2 \Omega_p} + \left(\frac{\mu}{\tau} + i k_y \frac{V_{gy}}{\Omega_p} + \frac{V_{gx}}{\Omega_p}  \frac{d}{d x} \right) \frac{\delta V_{gy}}{\Omega_p} = - i k_y \frac{\phi_p}{c_s^2} .\label{linear6}
\end{gather}
Here, $c_s$ represents the sound speed. $\mu$ denotes the dust-gas ratio, $f_d/f_g$. $\boldsymbol{W}_s \equiv \boldsymbol{V}_d - \boldsymbol{V}_g \equiv ( W_{s,x}, W_{s,y})$ denotes the relative velocity between dust and gas, and it will be called drift velocity hereafter. And 
\begin{gather}
    V_{dx} = - \frac{1}{2} f_g \chi_1 h_p \Omega_p ,\\
    V_{dy} = - \frac{3}{2} \Omega_p x - \frac{1}{2} f_g \chi_2 h_p \Omega_p, \\
    V_{gx} = \frac{1}{2} f_d \chi_1 h_p \Omega_p ,\\
    V_{gy} = - \frac{3}{2} \Omega_p x + \frac{1}{2} \left( f_d \chi_2 - 1 \right)h_p \Omega_p.
\end{gather}
$\phi_{p}$ is the planet potential, expressed by Bessel functions \citep{Goldreich1980,Rafikov2012}, i.e.
\begin{gather}
    \phi_p\left(k_y, x\right)=-\frac{G M_p}{\pi \mathrm{H}} K_0\left(\left|k_y \sqrt{x^2 + x_s^2} \right|\right), \\
    \frac{\partial \phi_p}{\partial x}=\operatorname{sgn}(x) \frac{G M_p}{\pi \mathrm{H}} k_y K_1\left(\left|k_y \sqrt{x^2 + x_s^2} \right|\right).
\end{gather}
$K_{n}$ is the modified Bessel function of order $n$. $x_s$ is the softening length, set to $1/8$ as used in \cite{Dong2011}. With the above preparations, we employ the relaxation method \citep{Huang2022,Press1992} to solve Equation \eqref{linear1}-\eqref{linear6}. This method addresses the two-point boundary value problem through an iterative process similar to the Newton-Raphson method. Equations \eqref{linear1}-\eqref{linear6} are discretized into finite difference equations (FDEs) within the computational domain. Both the PDEs and boundary conditions (BCs) are satisfied in each iteration. We implement free outgoing BCs with the WKBJ approximation, similar to the approach in \cite{Li2000}. To ensure sufficient convergence, we use a high resolution of $2\times10^{-4} \mathrm{H}$, achieving an error less than $10^{-9}$.

\section{Dusty \& Gaseous Modes} \label{results}
\subsection{Wavy Behaviors}
\begin{figure*}[htbp]
    \centering
    \includegraphics[width=1.0\textwidth]{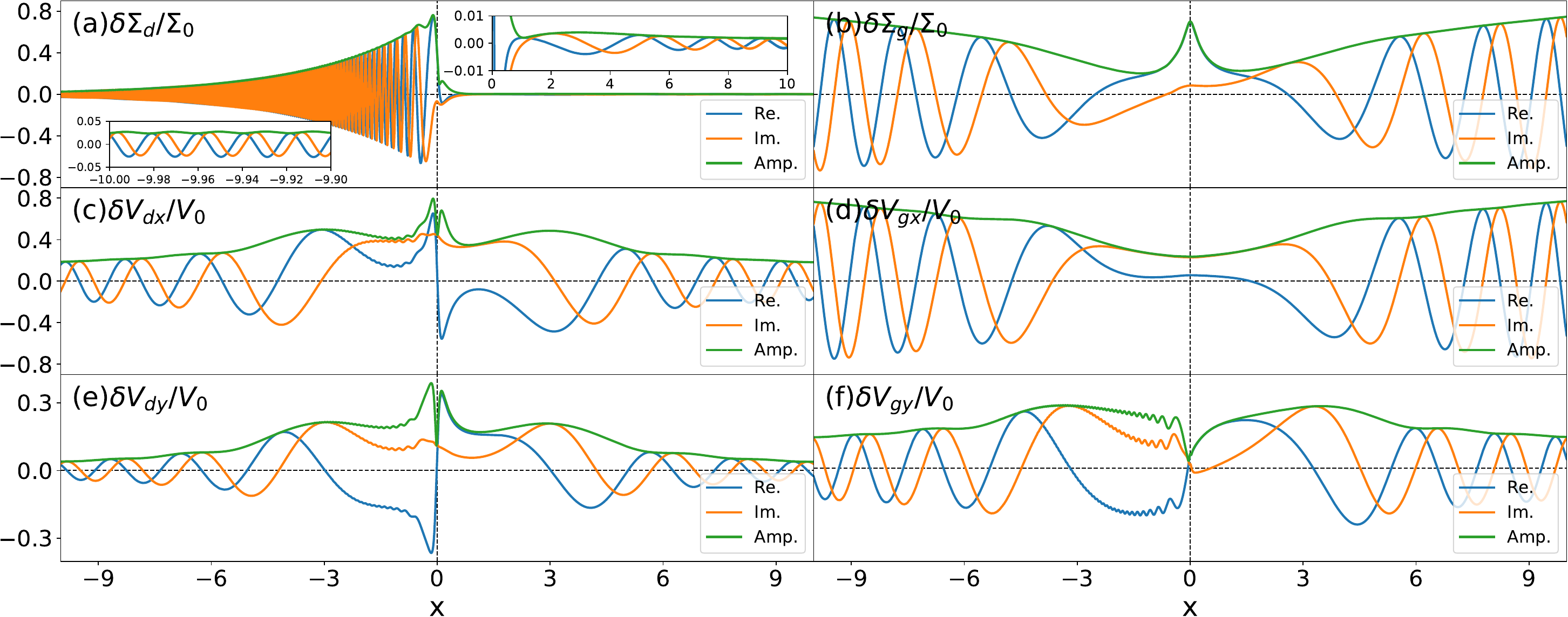}  
    \caption{Dusty and gaseous waves when $f_d=0.01,\tau=1.0,k_y=0.3$. All quantifies in the figure are normalized. (a) The dusty density perturbation. (b) The gaseous density perturbation. (c) The dusty radial velocity perturbation. (d) The gaseous radial velocity perturbation. (e) The dusty azimuthal velocity perturbation. (f) The gaseous azimuthal velocity perturbation.}
    \label{perturbations}
\end{figure*}

Previous studies by \cite{Squire2018a,Squire2018b} interpreted the YG streaming instability \citep{Youdin2005} as the resonance between epicyclic oscillations of gas and the drift velocity, termed as ``Resonant Drag Instability" (RDI). Furthermore, they suggest that the drift velocity can resonate with various waves, including sound waves \citep{Hopkins2018}. RDI occurs when the dust drift velocity exceeds the sound speed ($W_s > c_s$), leading to the emergence of instabilities. In our scenario, where $W_s \ll c_s$, RDI is absent. However, the planet can still excite certain modes.

Through solving \eqref{linear1}-\eqref{linear6} with relaxation method, we can get as many sets of solutions as the number of $k_y$. For a specific case with $k_y = 0.3$, Figure \ref{perturbations} illustrates the excitation of dusty and gaseous waves by a planet situated at the coordinate origin. And we select $f_d = 0.01$, $\tau = 1.0$ for a clear illustration. The perturbed densities and velocities are normalized by $\Sigma_{0} = q h_p^{-3}\Sigma_p$ and $V_0 = q h_p^{-3} \Omega_p$, where $q$ represents the mass ratio of the planet to the host star ($M_p/M_{\star}$). In the figure, we notice that such a dust mass fraction has a minimal impact on gas, as depicted in subfigures (b), (d), and (f). The results resemble those calculated in a pure gas disk \citep[e.g.][]{Tanaka2002,Rafikov2012}, except for the short-wavelength structure observed inside the planet. This short-wavelength structure arises due to the feedback of dust, leading us to refer to the gaseous waves as ``quasi-density" waves.

Dusty perturbations exhibit distinct characteristics compared to gaseous perturbations. Subfigures (a), (c), and (e) illustrate that dusty perturbations display a short-wavelength structure inside the planet. It originates from the ``quasi-drift'' mode \citep{Hopkins2018}. In next subsection, we will solve its radial wavenumber and demonstrate that the appearance of short-wavelength structure is robust. Identifying these features necessitates a high spatial resolution. The lower right zoom-in panel in subfigures (a) shows that the radial wavelength near the inner boundary is about $2\times10^{-2} \mathrm{H}$, which poses a significant challenge for resolution in hydrodynamic simulations. We use $2\times10^{-4} \mathrm{H}$ resolution, corresponding to $10^5$ mesh points, to ensure that the refined structure is presented in our calculations. With such a high resolution, the solutions become converged and have no dependency on the radial domain extent or mesh points. 

Away from the planet, dusty waves are dominated by ``quasi-density" waves, except for the density perturbation inside the planet as shown in subfigure (a). This dominance is due to the slow damping of the quasi-drift mode and the small amplitude of the dusty quasi-density wave. The latter is depicted in the upper right zoom-in panel, where $\delta \Sigma_{d} \sim f_d \delta \Sigma_{g}$. As $f_d$ increases, the quasi-drift mode is damped faster, and quasi-density waves also come to dominate wavy behaviors further inside the planet. This will be further illustrated in the next subsection.
\subsection{Dispersion Relation of the Quasi-drift Mode} \label{sec_dr}
\begin{figure}[htbp!]
    \centering
    \includegraphics[width=\columnwidth]{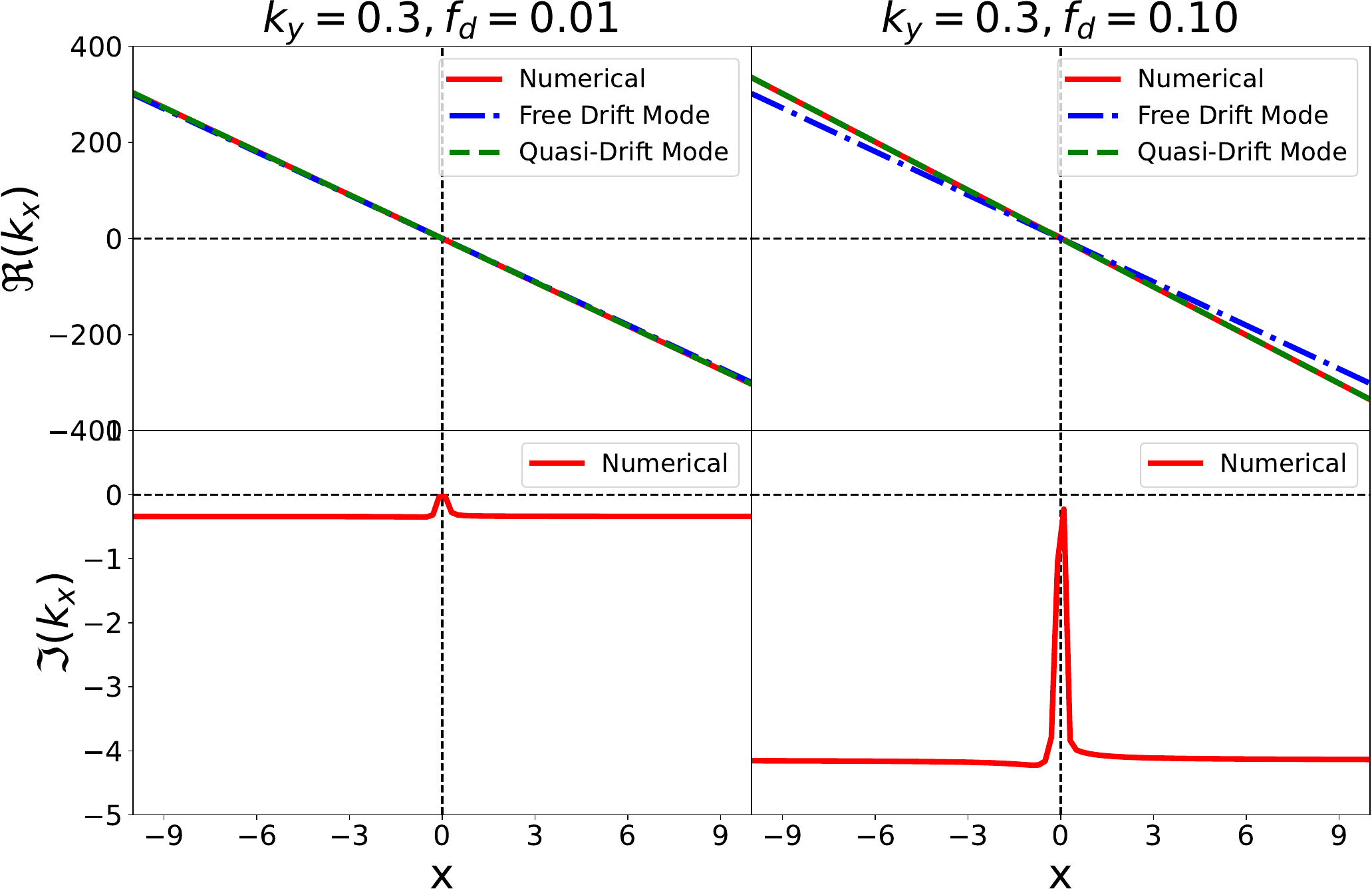}  
    \caption{The radial wavenumbers of quasi-drift modes. $\tau = 1.0$ and $k_y=0.3,f_d=0.01 \& f_d=0.1$ are compared in two columns. The upper(lower) row represents the real(imaginary) part. Red solid lines show our numerical results. Blue dot-dash lines show the radial wavenumber of free drift mode, i.e. $ k_x = \left( \tilde{\omega}_d - k_y W_{s,y} \right) / W_{s,x}$. Green dot-dash lines show the radial wavenumber of quasi-drift mode, i.e. $ k_x = \left( f_g \tilde{\omega}_d - k_y W_{s,y} \right) / W_{s,x}$.}
    \label{figwn}
\end{figure}

\cite{Hopkins2018} derived the dispersion relation for dust-gas systems in the free-falling frame. The free drift mode exhibits a dispersion relation of $ \omega = \boldsymbol{W}_s \cdot \boldsymbol{k}$. Similar results can be obtained in our framework. Employing WKB approximation, we numerically solve for the radial wavenumber of the quasi-drift mode. We obtained a fitting dispersion relation, $ \tilde{\omega} = \left( 1 + \mu \right) \boldsymbol{W}_s \cdot \boldsymbol{k}$. The results are depicted in Figure \ref{figwn}, where the upper(lower) panel displays the real(imaginary) part of the radial wavenumber, $k_x$. In the upper row, the red solid line represents our numerical results. The blue dot-dash line indicates $k_x$ of the free drift mode, i.e. $ k_x = \left( \tilde{\omega}_d - k_y W_{s,y} \right) / W_{s,x}$, where $\tilde{\omega}_d$ represents the Doppler-shifted frequency of dust, defined by $k_y r (\Omega_p - \Omega_d)$. The green dot-dash line indicates $k_x$ of the quasi-drift mode, i.e. $ k_x = \left( f_g \tilde{\omega}_d - k_y W_{s,y} \right) / W_{s,x}$. The quasi-drift mode aligns perfectly with our numerical calculations, indicating that these short-wavelength structures arise from this mode. Furthermore, $|k_x|$ increases linearly as $|x|$ increases. Specifically, at the inner boundary ($x = -10$), $k_x \sim 300$, a value consistent with the radial wavelength of $\sim 2\times10^{-2} \mathrm{H}$ as depicted in Figure \ref{perturbations}. Lastly, both the phase velocity and group velocity of the mode are $ \sim \boldsymbol{W}_s$, indicating that the wave propagates inward in the $-x$ direction and ahead of the planet in the $+y$ direction.

From the lower row, we find that the quasi-drift mode is excited by the planet and propagates inward while undergoing rapid damping, particularly at locations slightly further away from the planet. This behavior is consistent with the findings of \cite{Hopkins2018}. In the absence of planets, the mode remains stable due to $W_s \ll c_s$. Consequently, its structures dominate the region near the interior of the planet, displaying asymmetry in the shearing sheet. When $f_d = 0.01$, the damping rate is small, allowing the quasi-drift mode to dominate wavy behaviors far inside the planet. While $f_d > 0.01$, the damping rate increases rapidly away from the planet, causing quasi-density waves to dominate wavy behaviors, as illustrated in Figure \ref{comparison}.

\subsection{Planetary Wake}
\begin{figure*}[htbp!]
    \centering
    \includegraphics[width=\textwidth]{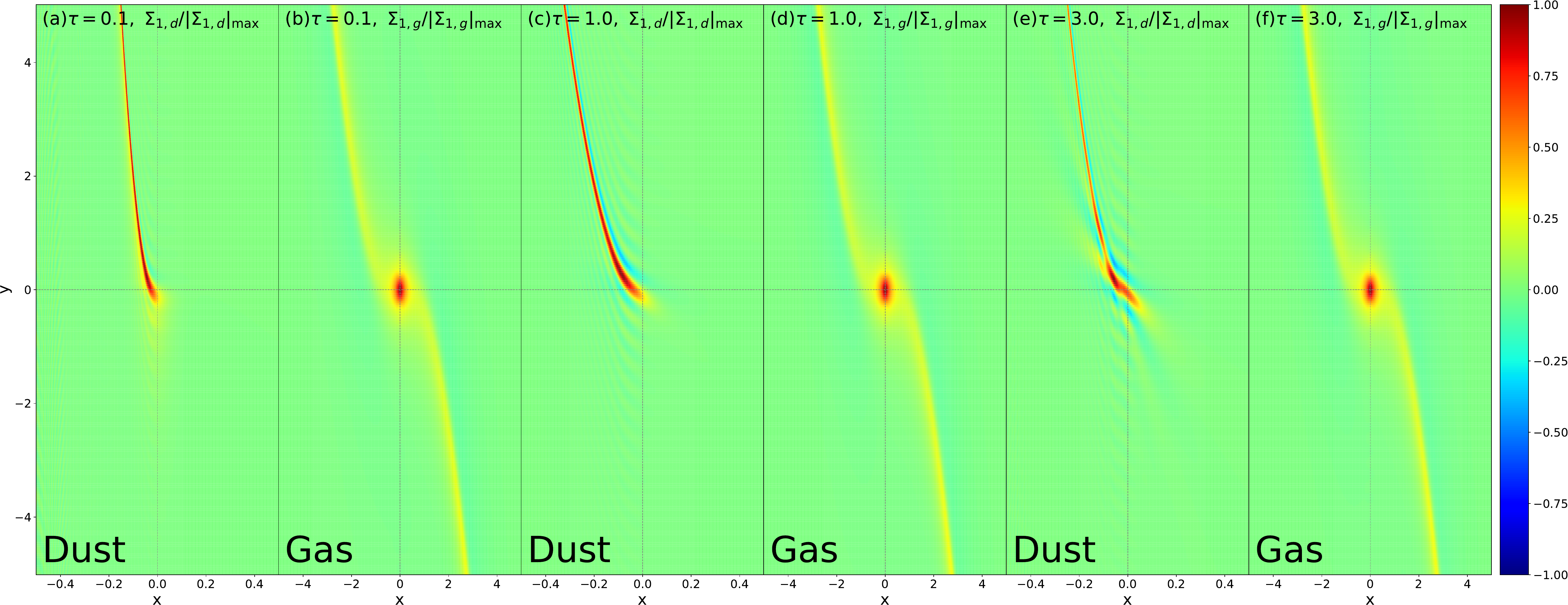}  
    \caption{Planetary wakes with $f_d = 0.01$. From left to right, $\tau = 0.1,1.0,3.0$, and every two show the dusty drift mode and gaseous density waves, separately (Note their ranges of $x$ axes are different). All perturbations in each panel are normalized by their respective maximum values.}
    \label{wake}
\end{figure*}

To visualize the waves in real space, we integrate density perturbations. Conclusively, the planetary wakes are depicted in Figure \ref{wake} with $f_d = 0.01$. From left to right, we consider $\tau = 0.1,1.0,3.0$, with every two panels displaying the dusty quasi-drift mode and gaseous quasi-density waves separately (note their ranges of $x$ axes differ). All perturbations in each panel are normalized by their respective maximum values. Comparing panels (b), (d), and (f), we find minimal changes in gaseous quasi-density waves. However, the dusty quasi-drift mode exhibits a short radial length-scale ($\sim 0.1 \mathrm{H}$). These results align with our previous analysis indicating that the quasi-drift mode is rapidly damped as it propagates inward. As we talked before, both the phase velocity and group velocity of the mode are $ \sim \boldsymbol{W}_s$. So, the quasi-drift mode remains ahead of the planet, exerting a positive torque on it, which is consistent with panels (a), (c) and (e). Due to the large amplitude of the mode, this effect is significant, termed as the ``Streaming Torque (ST)". When $\tau \sim 1$, the dusty wake exhibits a larger radial extent and relative density perturbation, indicating that the torque effect is most pronounced around $\tau \sim 1$, where the drift velocity, $\boldsymbol{W}_s$, reaches its maximum. Detailed calculations are presented in the subsequent section.

The morphologies of dust wave are consistent with \cite{Llambay2018}, who conducted hydrodynamic simulations for a dust-gas PPD. Their results present radial short wavelength structures inside the planet for as their figures 1 and 4 show. Dust surplus, deficit and multiple dust gaps inside the plane might imply the nonlinear evolution phase of dust drift mode.

\section{Disk Torque and Planetary Migration} \label{Torque}
\subsection{Angular Momentum Flux} \label{sec_amf}

\begin{figure*}[htbp!]
    \centering
    \includegraphics[width=\textwidth]{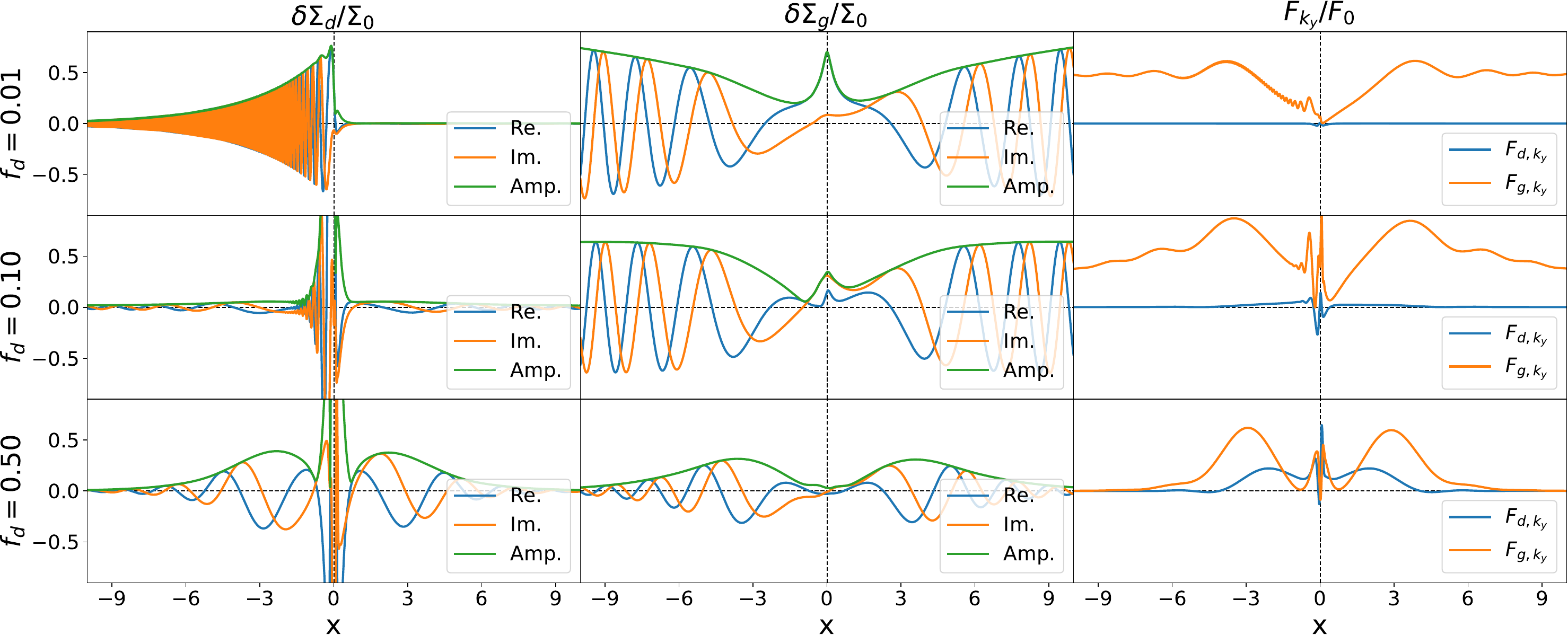}  
    \caption{Dusty and gaseous density perturbations and AMFs with $\tau = 1.0$, $k_y=0.3$, and varying $f_d$ each row. With $f_d$ increasing, both dusty and gaseous AMFs diminish progressively, indicative of damped ``quasi-density" waves, and the quasi-drift mode experiences heightened damping.}
    \label{comparison}
\end{figure*}

For low-mass planets embedded within pure gas PPDs, the Lindblad torque (LT), $\Gamma_{\mathrm{L}}$, and corotation torque (CT), $\Gamma_{\mathrm{C}}$, exerted by the disk on the planet play crucial roles in driving type I migration \citep{Ward1986,Goldreich1979,Korycansky1993,Tanaka2002}. These torques can be quantified by analyzing the AMF carried by density waves, given by:

\begin{gather}
    F = \int_{0}^{\infty} F_{k_y} dk_y, \label{amf} \\
    F_{k_y}= 4 \pi r_p^2 \Sigma_p \left[ \Re\left(\delta V_x\right) \Re\left(\delta V_y\right)+\Im\left(\delta V_x\right) \Im\left(\delta V_y\right) \right] . \label{amf_ky}
\end{gather}
Then LT and CT,  are given by
\begin{gather}
    \Gamma_{\mathrm{L/C}, k_y} = \int_{0}^{\infty} \Gamma_{\mathrm{L/C}, k_y} dk_y, \\
    \Gamma_{\mathrm{L}, k_y}=F_{k_y}(x \rightarrow - \infty)-F_{k_y}(x \rightarrow+\infty), \\
    \Gamma_{\mathrm{C}, k_y}= F_{k_y}(x \rightarrow+0)-F_{k_y}(x \rightarrow-0). \label{torque_c}
\end{gather}
For a two-fluid system, distinct AMFs arise for dusty and gaseous waves. Figure \ref{comparison} illustrates the density perturbations and AMFs for both components with $\tau = 1.0$, $k_y=0.3$, and varying $f_d$. And the normalization is $F_{0} = q^2 h_p^{-3} \Sigma_{p} r_p^4 \Omega_{p}^2$. The first row, consistent with Figure \ref{perturbations}, demonstrates minimal dusty AMF due to a low $f_d$. Gaseous AMF tends towards a constant at large distances from the planet. However, when $f_d$ increases, as shown in subsequent rows, both dusty and gaseous AMFs diminish progressively, indicative of damped ``quasi-density" waves.It means we cannot get the disk torque on planets through \eqref{amf}-\eqref{torque_c}. Actually, Equation \eqref{amf_ky} does not apply in dissipation flows \citep{Balbus2003}. In the system, dust and gas interacts through friction. Then, part of kinetic energy is dissipated into heating. But we use isothermal state for gas. The cooling process would exert a thermal damping on propagating waves. Similar behaviors appear in \cite{Miranda2020} because of thermal damping. However, our current investigation focuses on low $f_d$ and the quasi-drift mode, which exhibits substantial torques. So, we do not go into detail about the damped AMFs. Notably, the quasi-drift mode experiences heightened damping as $f_d$ increases, which is consistent with the discussion in section \ref{sec_dr}. Ultimately, we refrain from utilizing Equations \eqref{amf} to \eqref{torque_c} for torque calculations. This decision stems from the inherent limitation of this method in capturing ST effects accurately.

\subsection{Torque Calculations}
We can calculate the torque by integrating planetary force, which reads
\begin{equation}
    \Gamma= r_p \int_{-\infty}^{\infty} \int_{-\infty}^{\infty} dx d y \Sigma_1 \frac{\partial \phi_p}{\partial y}.
\end{equation}
In Fourier space,
\begin{gather} \label{torque_eq}
    \Gamma=  \int_0^{\infty} \int_{-\infty}^{\infty} \left(\frac{d \Gamma}{d x}\right)_{k_y} dx d k_y, \\
    \left(\frac{d \Gamma}{d x}\right)_{k_y}= -4 \pi r_p k_y \phi \Im(\delta \Sigma) .
\end{gather}
We use Equation \eqref{torque_eq} with $k_y$ ranging from $0.01$ to $15$, logarithmically discrete to 320 points. The convergence for total torques is confirmed with high spatial resolution.

\begin{figure}[htbp!]
    \centering
    \includegraphics[width=\columnwidth]{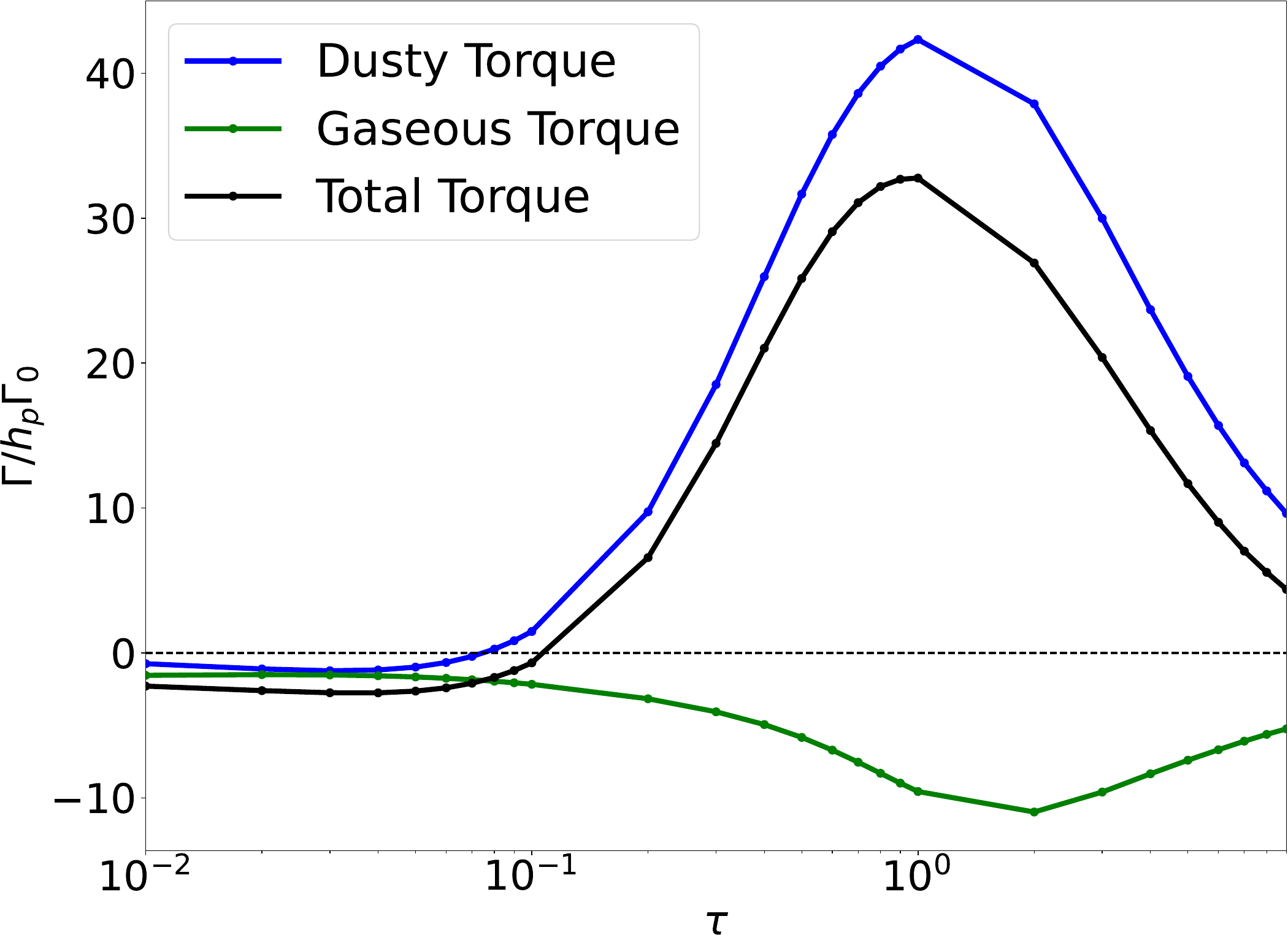}  
    \caption{The disk torques on planets with $f_d = 0.01$ and different $\tau$ values. Solid dots are from our calculations, linked by three curves. The blue curve shows torques from dust modes. The green curve show those from gas modes. And the black curve shows the total torque.}
    \label{torque}
\end{figure}

The results are presented in Figure \ref{torque} for $f_d=0.01$. The $x$-axis represents different $\tau$ values. The solid dots represent our calculations, linked by three curves. The blue curve indicates torques from dusty modes, labeled as ``Dusty Torque", while the green curve represents those from gaseous modes, labeled as ``Gaseous Torque". The black curve depicts the total torque. Here, $h_p \Gamma_0 = q^2 h_p^{-2} \Sigma_{p} r_p^4 \Omega_{p}^2$. We set $h_p = 0.03$ for comparison with \cite{Tanaka2002}.

The torque analysis in Figure \ref{torque} reveals several key insights. Dusty torques predominantly exhibit positive and dominate total torques in most scenarios. Conversely, gaseous torques appear negative, driving inward migration for planets solely.

Dusty torques reach extreme points when $\tau \sim 1$, corresponding to the peak drift velocity. The dominance of dusty torques is attributed to the ST, which surpasses contributions from CT and LT for dust. Notably, when $\tau$ is sufficiently low ($\lesssim 10^{-1}$), dusty torques become negative due to strong gas-dust coupling, causing dust behavior to mimic that of gas.

The negative nature of gaseous torques aligns with their role in facilitating inward migration, a phenomenon extensively studied in prior literature \citep[e.g.][]{Ward1986,Ward1997,Korycansky1993,Tanaka2002}. If we consider $\tau$ to be very low or large, the gaseous torque must converge to that in a pure gas PPD. \cite{Tanaka2002} employed the ``Modified Local Approximation", which incorporates the curvature effect, surface density gradient, and pressure gradient compared to the shearing sheet. In this paper, only the effect of the pressure gradient is considered, resulting in a change of the rotation curve from Keplerian frequency. This alteration affects the locations of Lindblad resonance (LR) points, leading to a net LT. Our calculations yield $\Gamma_g \simeq - 2.0 h_p \Gamma_0$, which is consistent with the value obtained by \cite{Tanaka2002} of $\Gamma_g \simeq - 2.3 h_p \Gamma_0$, attributed solely to the pressure gradient. When $\tau \sim 1$, the gaseous radial velocity amplifies CT, leading to maximum negative gaseous torque. This phenomenon is reminiscent of type III migration, originating from planetary migration \citep{Masset2003, Ogilvie2006, Paardekooper2014}.

\subsection{Planetary Migration}
It is then easy to estimate the planetary migration timescale. Using some typical parameters, \cite{Tanaka2002} found it would be typically $10^{6}$ years with inward migration, which is comparable to or shorter than the lifetime of PPDs. \cite{Tanaka2002} gives the (outward) migration timescale
\begin{equation}
    \tau_{\mathrm{mig}} = \frac{L_p}{2\Gamma},
\end{equation}
where $L_p$ is the angular momentum of the planet $M_p(GM_{\star}r_p)^{1/2}$. With $f_d = 0.01$, Figure \ref{torque} allows $\Gamma$ ranges from $-3h_p \Gamma_0$ to $30 h_p \Gamma_0$. Using the same parameters as those used in \cite{Tanaka2002}, the final result would be $ - 10^6$ to $10^5$ years (the negative sign means inward migration), depending on $\tau$. This illustrates how the ST can dominate both the CT and LT, consequently driving outward planetary migration.

\section{Discussions and Conclusions} \label{Conclusions}

In this paper, we explore planetary migration in dust-gas coupled PPDs. Employing linear calculations with NSH equilibrium under the shearing sheet approximation, we unveil the significant contribution of the dusty quasi-drift mode to planetary migration.

In such two-fluid systems, the quasi-drift mode emerges and governs the wavy behaviors in close proximity to planets. As the dust fraction increases, the amplitude of dusty quasi-density waves grows, while the quasi-drift modes are damped rapidly. Consequently, quasi-density waves, unable to propagate within the corotation region, take precedence in regions farther from the planet. The quasi-drift mode features extremely short radial wavelength, with a dispersion relation of $ \tilde{\omega} = \left( 1 + \mu \right) \boldsymbol{W}_s \cdot \boldsymbol{k}$, indicating its subsonic nature, originating from dust drift.

The quasi-drift mode engenders a wake characterized by a short radial length-scale, positioned ahead of the planet, yielding a positive torque, referred to ``Streaming Torque''. Additionally, the NSH velocity alters both LT and CT, with LR points undergoing a shift due to sub-Keplerian rotation and CT experiencing amplification from radial velocity.

For a typical dust fraction ($\sim 0.01$), the final outcome of planetary migration depends on the degree of coupling between dust and gas, quantified by the dimensionless stopping time, $\tau$. We find that at low $\tau$ values, both dust and gas contribute negative torque, resulting in inward migration. Conversely, when $\tau$ approaches unity, both dusty and gaseous torques peak. Notably, the dominance of dusty ST, originating from the quasi-drift mode, drives a rapid outward migration with characteristic timescales $10^5\mathrm{yr}$, which can solve the existing puzzle that planets migrate inward fast. We also conduct a tentative survey for the torques with different $f_d$. Near $f_d = 0.01$, we find the absolute ST vaule increases with higher $f_d$. It means planetary migration will be faster with a higher $f_d$. At the moment, nonlinear effects and other physical ingredients, like planetesimal formation, might be important. Then, the ST calculations with higher $f_d$ should be calculated with more consideraitons.

An essential concern is the softening length, with a value of $1/8$ utilized in this work, as validated in \cite{Dong2011}. Furthermore, the consistency of gaseous torques with \cite{Tanaka2002} is crucial for the quantitative calculation of ST. We notice that the quasi-drift mode is excited near the planet and sensitive to planetary potential, which means it must be careful with the use of softening length.

Our findings are consistent with \cite{Llambay2018}, which conducted hydrodynamic simulations for dust-gas PPDs. They found that when $\tau$ is low, dusty torque is negative, while at higher $\tau$ values, a dusty cavity behind the planet results in a large positive torque. Radial short wavelength structures inside the planet of dust also appear in their simulations as their figures 1 and 4 show, which imply the nonlinear phase of our calculations.

Exploring the vast parameter space, including dust mass fraction and aspect ratio, is crucial. While standard PPDs are presumed to have $f_d \sim 0.01$, recent findings by \cite{Stefansson2023} suggest it could be underestimated, possibly reaching about $0.10$. We also do not consider an Epstein regime, unlike \cite{Hopkins2018} allowing a variation for stopping time. But there is no fundamental physical difference between them and ST still works. In this work, we assume the PPD as a thin disk and use vertically integrated equations, which is used widely for the investigation of planetary migration \citep[e.g.][]{Goldreich1980,Korycansky1993}. However, the effects of vertical structure on ST deserves future identification \citep{Tanaka2002,Tanaka2024}. Other details and more physical ingredients, such as dust diffusion and accretion, also warrants further investigation. Dust diffusion tends to restrain the waves with short scale, which means it might weaken the effect of ST, similar to the stabilization of streaming instability \citep{Chen2020,Umurhan2020}. It has been found that gas accretion contribute a lot to planetary migration \citep{Li2024,Laune2024}. For a low-mass planet, dust accretion might affect the ST through changing the drift velocities nearby. We will investigate both of them in the future.

In conclusion, the relative motion of dust and gas facilitates the existence of the dusty drift mode, resulting in ST and altering the migration history of low-mass protoplanets in the early phase. Depending on the coupling between dust and gas, outward migration becomes a viable scenario.

\section*{Acknowledgement}
We thank the anonymous referee for the useful comments and suggestions that improve the manuscript. Q.H. thank Shunquan Huang for the fruitful discussion. This work has been supported by the National SKA Program of China (Grant No. 2022SKA0120101) and National Key R\&D Program of China (No. 2020YFC2201200) and the science research grants from the China Manned Space Project (No. CMS-CSST-2021-B09, CMS-CSST-2021-B12, and CMSCSST-2021-A10) and opening fund of State Key Laboratory of Lunar and Planetary Sciences (Macau University of Science and Technology) (Macau FDCT Grant No. SKL-LPS(MUST)-2021-2023). C.Y. has been supported by the National Natural Science Foundation of China (grants 
11521303, 11733010, 11873103, and 12373071).

\vspace{5mm}
\bibliography{dust_gas.bib}{}
\bibliographystyle{aasjournal}
\end{document}